\documentclass[aps,twocolumn]{revtex4-1}
\usepackage{amsmath}
\usepackage{graphicx}

\begin{document}

\title{Off-diagonal long-range order in the ground state of the Kitaev chain 
}
\author{E. S. Ma}
\author{Z. Song}
\email{songtc@nankai.edu.cn}
\affiliation{School of Physics, Nankai University, Tianjin 300071, China}

\begin{abstract}
We study a one-dimensional Kitaev model with uniform phase gradient pairing
term. We show that the gradient constant dramatically affects the phase
diagram, which consists of topologically trivial and nontrivial phases,
associated with Majorana edge modes. Based on the exact solution, a
Bardeen-Cooper-Schrieffer (BCS)-pair order parameter is introduced to
characterize the phase diagram by its value and nonanalytic behavior at
phase boundaries. We find that this order parameter obtains its maxima at
the triple critical points, at which the pairing phase gradient suppresses
the single-particle scattering process due to the coherent destructive
interference. In particular, we show that the ground state at such a point
possesses exact off-diagonal long-range order (ODLRO), in the thermodynamic
limit. Our result provides an example of a gapless $p$-wave superconducting
ground state possessing ODLRO.
\end{abstract}

\maketitle

\affiliation{School of Physics, Nankai University, Tianjin 300071, China}

\affiliation{School of Physics, Nankai University, Tianjin 300071, China}

\affiliation{School of Physics, Nankai University, Tianjin 300071, China}

\section{Introduction}

The one-dimensional (1D) Kitaev model \cite{Kitaev} is a prototype minimal
model of interacting spinless fermions, describing $p$-wave topological
superconductors \cite{qi2011zhang}. It hosts Majorana zero modes \cite%
{Sarma,Stern,Alicea}, which have attracted tremendous attention in condensed
matter and materials physics communities due to the implications in
topological quantum computation. This model consists of three terms:
spin-less fermion hopping between lattice sites, chemical potential and pair
creation (annihilation) on a dimer. The simplicity of the model permits the
exact solution, which exhibits a rich variety of emergent quantum many-body
phenomena such as superconductivity and Majorana zero modes. Unlike another
spinful interacting fermionic model, the Hubbard model, the ground state of
the Kitaev model can be a superconducting state. In parallel, although the
Hubbard model also involves the terms of hopping and on-site pair
interaction, the ground state is not a superconducting state. However, it
permits a set of special exact eigenstates, $\eta $-pairing states, which
describe the\ condensation of on-site fermion pairs with the identical
momentum $\pi $. Remarkably, these states are shown exactly to possess
off-diagonal long-range order (ODLRO) \cite{yang1989eta, yang1990so}. Exact
expression of a many-fermion state has a unique advantage to identify the
features of ODLRO. A natural question arises from the existence of ODLRO in
the ground state of the Kitaev model. Understanding the properties of the
Kitaev model provides insights into how order can emerge from the interplay
among the single-particle kinetic energy, chemical potential, and pairing
amplitude.

In this work, we consider the 1D Kitaev model with an additional parameter,
phase on the strength of the pairing term with a uniform gradient. We show
that the gradient constant dramatically affects the phase diagram, which
consists of topologically trivial and nontrivial phases, associated with
Majorana edge modes as a demonstration of bulk-boundary correspondence
(BBC). Specifically, we examine the pairing mechanism by introducing a
Bardeen-Cooper-Schrieffer (BCS)-pair order parameter. Based on the exact
solution, we find that the phase diagram can be used to characterize the
value of the parameter and its nonanalytic behavior at phase boundaries.
This order parameter obtains its maxima at the triple critical points, at
which the pairing phase gradient suppresses the single-particle scattering
process due to the coherent destructive interference. Such a ground state is
essentially the\ condensation of fermion pairs with zero momentum, which is
also shown to be equivalent to that of spatially local fermion pairs. As
expected, this condensation results in the existence of an exact ODLRO in
the thermodynamic limit.\ Our result provides an example of a gapless $p$%
-wave superconducting ground state possessing ODLRO.

The rest of the paper is organized as follows. Sec. \ref{Model and phase
diagram} discusses the Kitaev model, wherein the exact solution for periodic
and open boundary conditions are obtained and the phase diagram is
presented. Sec. \ref{Order parameter} introduces the BCS-like parameter for
the analytical understanding of the $p$-wave pairing ground state and the
pair condensation. Sec. \ref{Pair condensation with ODLRO} shows the
existence of ODLRO for the condensation of zero-momentum fermion pairs. Sec. %
\ref{Summary} concludes this paper. Some details of our calculations are
placed in the Appendix.

\begin{figure}[tbh]
\centering
\includegraphics[width=0.5\textwidth]{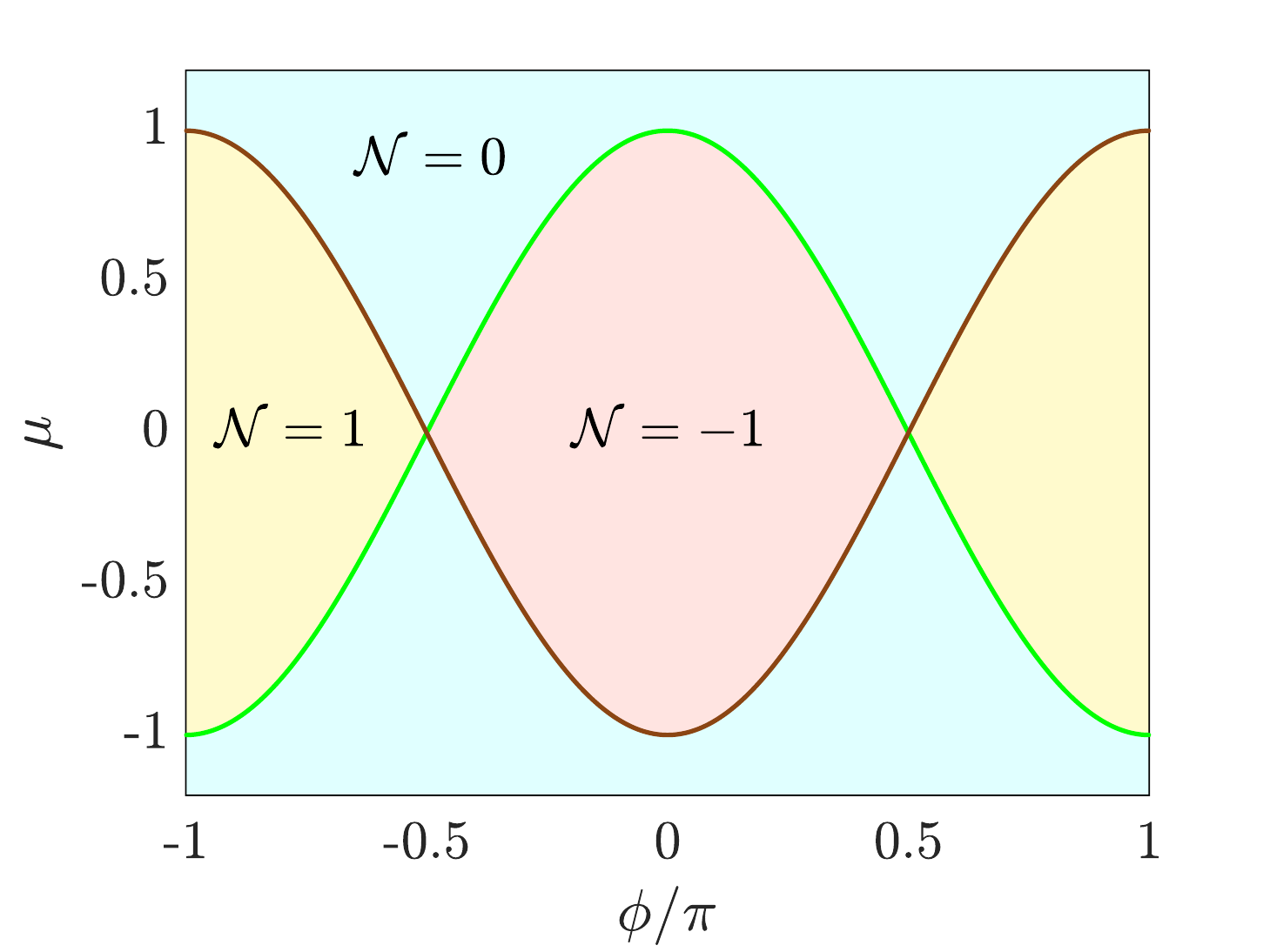}
\caption{Phase diagram of the Hamiltonian in Eq. (\protect\ref{H1}) on the
parameter $\protect\phi -\protect\mu $ plane. Different color regions
represent different phases that are distinguished by winding number $%
\mathcal{N}$, which is defined in Eq. (\protect\ref{winding N}). The brown
and green lines are the boundaries of different phases, from Eq. (\protect
\ref{boundary}). This indicates that the $\protect\phi $\ factor shrinks the
region of the nontrivial phase in the chemical potential, resulting in
triple critical points at ($\pm \protect\pi /2$, $0$).}
\label{fig1}
\end{figure}

\section{Model and phase diagram}

\label{Model and phase diagram}

We begin with a Kitaev model with a phase gradient in the pairing terms by
considering the following fermionic Hamiltonian on a lattice of length $N$

\begin{eqnarray}
H &=&\sum\limits_{j=1}^{N}[-Jc_{j}^{\dag }c_{j+1}-\Delta e^{i\theta
_{j}}c_{j}^{\dag }c_{j+1}^{\dag }+\mathrm{H.c.}  \notag \\
&&+\mu \left( 2n_{j}-1\right) ],  \label{H}
\end{eqnarray}%
where $c_{l}^{\dag }$ $(c_{l})$\ is a fermionic creation (annihilation)
operator on site $l$, $n_{l}=c_{l}^{\dag }c_{l}$, $J$ is the tunneling rate, 
$%
\mu
$ is the chemical potential, and position-dependent complex strength $\Delta
e^{i\theta _{j}}$\ of the $p$-wave pair creation (annihilation). For a
closed chain, we define $c_{N+1}=c_{1}$, and for an open chain, we set $%
c_{N+1}=0$. The Kitaev model is known to have a rich phase diagram in its
simplest version, i.e., $\theta _{j}=0$. It has been shown that the
nonzero $\theta _{j}$\ affects the phase diagram and Majorana edge modes 
\cite%
{seradjeh2011unpaired,romito2012manipulating,rontynen2014tuning,dmytruk2019majorana,takasan2022supercurrent}%
. In this work, we only consider the Hamiltonian with a uniform phase
gradient, $\theta _{j}=2j\phi $ and $\Delta >0$. Mathematically, it can be
reduced to a familiar form%
\begin{eqnarray}
H &=&\sum\limits_{j=1}^{N}[-Je^{i\phi }c_{j}^{\dag }c_{j+1}-\Delta
c_{j}^{\dag }c_{j+1}^{\dag }+\mathrm{H.c.}  \notag \\
&&+\mu \left( 2n_{j}-1\right) ].  \label{H1}
\end{eqnarray}%
for the sake of convenience of discussion by taking the guage transformation 
$c_{j}\longrightarrow e^{i\left( j-\frac{1}{2}\right) \phi }c_{j}$. In the
following, we focus on the above Hamiltonian, considering the solutions of $%
H $\ under periodic and open boundary conditions. The obtained result can be
mapped to that of the original Hamiltonian.

The Hamiltonian is exactly solvable due to the translational symmetry of the
system, i.e., $\left[ T,H\right] =0$,\ where $T$\ is the translational
operator, defined by $Tc_{l}T^{-1}=c_{l+1}$. Taking the Fourier
transformation

\begin{equation}
c_{j}=\frac{1}{\sqrt{N}}\sum\limits_{k}e^{ikj}c_{k},
\end{equation}%
with $k=2\pi l/N$ ($l\in Z$) for the Hamiltonian (\ref{H1}), we have its
Nambu representation $H=\sum_{\pi >k>0}H_{k}$ with 
\begin{eqnarray}
H_{k} &=&-2\left( 
\begin{array}{cc}
c_{k}^{\dag } & c_{-k}%
\end{array}%
\right) h_{k}\left( 
\begin{array}{c}
c_{k} \\ 
c_{-k}^{\dag }%
\end{array}%
\right)  \notag \\
&&-2\cos (-k+\phi ),
\end{eqnarray}%
where the matrix 
\begin{equation}
h_{k}=\sum_{\alpha =0}^{3}B_{\alpha }\sigma _{\alpha },
\end{equation}%
is expressed by the Pauli matrices%
\begin{eqnarray}
\sigma _{0} &=&\left( 
\begin{array}{cc}
1 & 0 \\ 
0 & 1%
\end{array}%
\right) ,\sigma _{1}=\left( 
\begin{array}{cc}
0 & 1 \\ 
1 & 0%
\end{array}%
\right) ,  \notag \\
\sigma _{2} &=&\left( 
\begin{array}{cc}
0 & -i \\ 
i & 0%
\end{array}%
\right) ,\sigma _{3}=\left( 
\begin{array}{cc}
1 & 0 \\ 
0 & -1%
\end{array}%
\right) ,
\end{eqnarray}%
and the field components are%
\begin{eqnarray}
B_{0} &=&-\sin \phi \sin k,B_{1}=0,  \notag \\
B_{2} &=&-\Delta \sin k,B_{3}=\cos \phi \cos k-\mu .
\end{eqnarray}%
Here we take $J=1$ for the convenience of further analysis and the ($%
c_{k},c_{k^{\prime }}^{\dag }$)\ operators are fermion operators satisfying
commutation relations $\left\{ c_{k},c_{k^{\prime }}^{\dag }\right\} =\delta
_{kk^{\prime }}$ and $\left\{ c_{k},c_{k^{\prime }}\right\} =0$.

The eigenvectors of $h_{k}$\ are

\begin{equation}
\left\vert \psi _{k}^{\pm }\right\rangle =\left( 
\begin{array}{c}
\cos \frac{\theta _{k}}{2} \\ 
\sin \frac{\theta _{k}}{2}e^{i\varphi }%
\end{array}%
\right) ,\left( 
\begin{array}{c}
\sin \frac{\theta _{k}}{2} \\ 
-\cos \frac{\theta _{k}}{2}e^{i\varphi }%
\end{array}%
\right) .
\end{equation}%
with eigenvalues 
\begin{equation}
\varepsilon _{k}=\pm \sqrt{B_{2}^{2}+B_{3}^{2}}+B_{0}=\pm r_{k}-\sin \phi
\sin k,  \label{ek}
\end{equation}%
where the norm of Bloch vector $\mathbf{B}=(B_{1},B_{2},B_{3})$ is 
\begin{equation}
r_{k}=\sqrt{\left( \cos \phi \cos k-\mu \right) ^{2}+\left( \Delta \sin
k\right) ^{2}}.
\end{equation}
The angles read 
\begin{equation}
\sin \varphi =-\mathrm{sign}\left( k\right) ,\theta _{k}=\tan ^{-1}\frac{%
\Delta \left\vert \sin k\right\vert }{\cos \phi \cos k-\mu }.  \label{theta}
\end{equation}%
The diagonalizable form of the Hamiltonian is

\begin{equation}
H=\sum_{\pi >k>-\pi }[2\left( \sin \phi \sin k-r_{k}\right) \gamma
_{k}^{\dag }\gamma _{k}+r_{k}],
\end{equation}%
where the fermion operator is 
\begin{equation}
\gamma _{k}=\cos \frac{\theta _{k}}{2}c_{k}+\sin \frac{\theta _{k}}{2}%
e^{-i\varphi }c_{-k}^{\dag }.
\end{equation}

In this work, we only focus on the case with $\Delta >J>0$, in which the
ground state has the form

\begin{equation}
\left\vert G\left( \mu ,\phi \right) \right\rangle =\prod\limits_{\pi
>k>0}\left( \sin \frac{\theta _{k}}{2}-i\cos \frac{\theta _{k}}{2}%
c_{-k}^{\dag }c_{k}^{\dag }\right) \left\vert 0\right\rangle _{k}\left\vert
0\right\rangle _{-k},
\end{equation}%
with the density of the ground state energy

\begin{equation}
\mathcal{E}_{\mathrm{g}}=-\frac{2}{N}\sum_{\pi >k>0}r_{k}+\frac{\left\vert
\cos \phi -\mu \right\vert +\left\vert \cos \phi +\mu \right\vert }{N},
\end{equation}%
for even\textbf{\ }$N$\textbf{.}

The phase diagram can be determined by the condition%
\begin{equation}
r_{k_{c}}=0,
\end{equation}%
which results in $k_{c}=0$ or $\pi $, and the phase boundary%
\begin{equation}
\mu =\pm \cos \phi .  \label{boundary}
\end{equation}%
The topological index, winding number or Zak phase can be extracted from the
Bloch vector $\mathbf{B}=(B_{1},B_{2},B_{3})$. The winding number of a
closed curve in the auxiliary $B_{3}B_{2}$-plane around the origin\textbf{\ }%
is given by 
\begin{equation}
\mathcal{N}=\frac{1}{2\pi }\oint_{C}\mathbf{(}\hat{B}_{3}\mathrm{d}\hat{B}%
_{2}-\hat{B}_{2}\mathrm{d}\hat{B}_{3}),  \label{winding N}
\end{equation}%
where the unit vector $\mathbf{\hat{B}}\left( k\right) =\mathbf{B}\left(
k\right) /\left\vert \mathbf{B}\left( k\right) \right\vert $. $\mathcal{N}$
is an integer representing the total number of times that a curve travels
counterclockwise around the origin. Actually, the winding number is simply
related to the loop described by equation

\begin{equation}
\frac{(B_{3}+\mu )^{2}}{\cos ^{2}\phi }+\frac{\left( B_{2}\right) ^{2}}{%
\Delta ^{2}}\mathbf{=}1,  \label{elllipse}
\end{equation}%
which presents a normal ellipse in the $B_{3}B_{2}$-plane. In Fig. 1, the
winding number $\mathcal{N}$\ in each phase is given. In Fig. 2, the graphs
of the Bloch vector\ in (\ref{elllipse}) with winding number $\mathcal{N}$\
in each region is given in comparison with the spectrum of the Majorana
lattice.

\begin{figure}[tbh]
\centering
\includegraphics[width=0.5\textwidth]{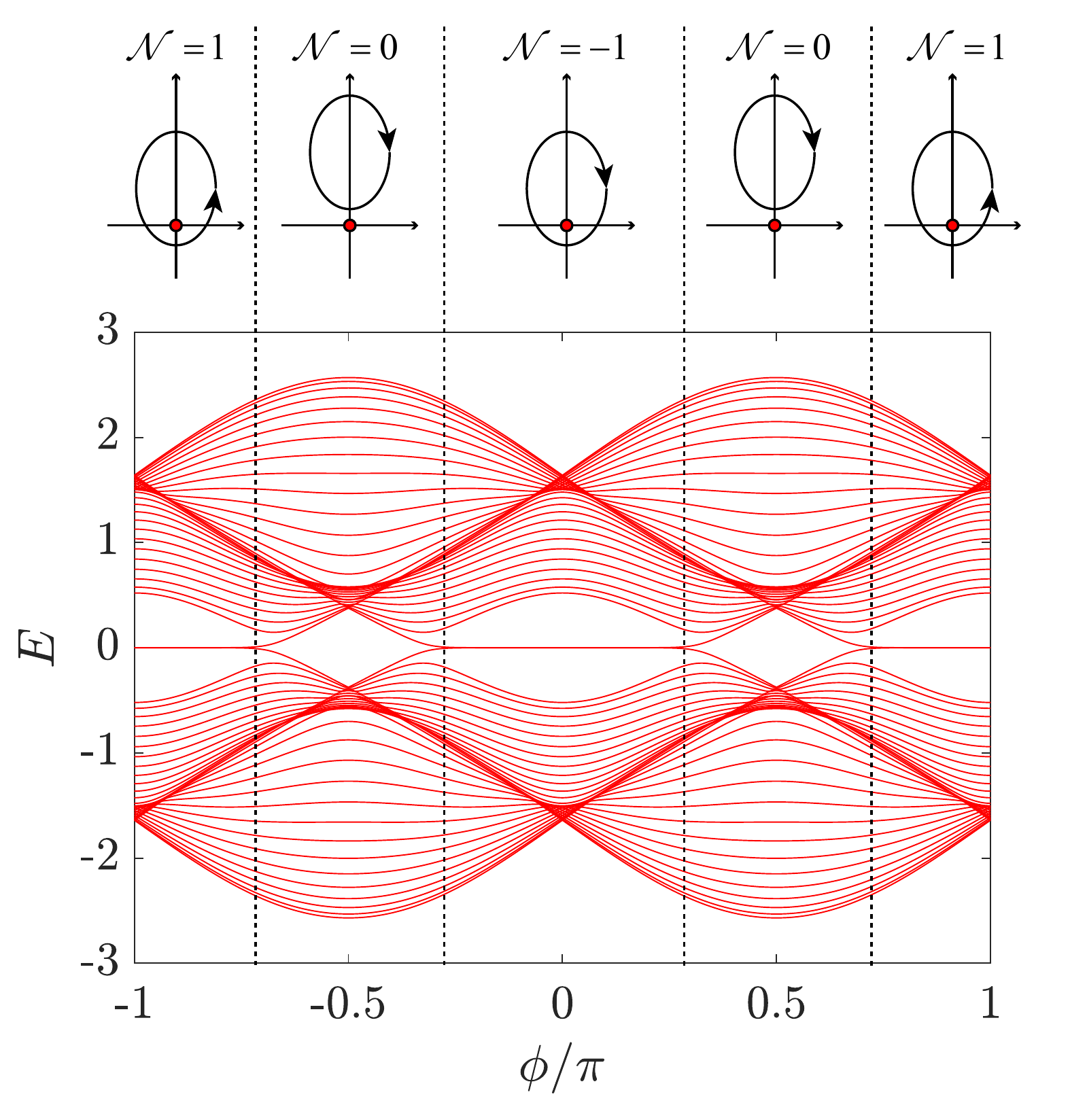}
\caption{Plots of the energy spectrum of the Majorana lattice in (\protect
\ref{Hm}) under open boundary condition obtained from numerical
diagonalization. The parameters are $\protect\mu =0.5$, $J=1$, $\Delta =1.5$
and $N=30$. The graphs of Bloch vector $\mathbf{B}$ in each region of $%
\protect\phi $ illustrate the corresponding winding numbers. Red filled
circles indicate the origin. We see that the mid-gap zero modes appear when
the winding number is nonzero, demonstrating the bulk-boundary
correspondence.}
\label{fig2}
\end{figure}

For a Kitaev chain, the translational symmetry is broken, and the above
solution is invalid. To obtain the solution of $H$ with an open boundary
condition, we introduce Majorana fermion operators

\begin{equation}
c_{j}=\frac{a_{j}-ib_{j}}{2},c_{j}^{\dag }=\frac{a_{j}+ib_{j}}{2},
\end{equation}%
which obey the commutation relations

\begin{equation}
\left\{ a_{j},a_{j^{\prime }}\right\} =2\delta _{jj^{\prime }},\left\{
b_{j},b_{j^{\prime }}\right\} =2\delta _{jj^{\prime }},\left\{
a_{j},b_{j^{\prime }}\right\} =0.
\end{equation}%
Then, the Hamiltonian can be rewritten in the form

\begin{equation*}
H=i\varphi ^{\mathrm{T}}h\varphi ,
\end{equation*}%
where the operator vector is

\begin{equation}
\varphi =\left( a_{1},b_{1},a_{2},b_{2},\cdots a_{N},b_{N}\right) .
\end{equation}%
Here $h$ is a $2N\times 2N$ matrix%
\begin{eqnarray}
&&h=\frac{1}{4}\sum_{j=1}^{N}[\left( \cos \phi -\Delta \right) \left\vert
j\right\rangle _{AB}\left\langle j+1\right\vert  \notag \\
&&-(\cos \phi +\Delta )\left\vert j\right\rangle _{BA}\left\langle
j+1\right\vert -2\mu \left\vert j\right\rangle _{AB}\left\langle
j\right\vert ]  \notag \\
&&-\sin \phi (\left\vert j\right\rangle _{AA}\left\langle j+1\right\vert
+\left\vert j\right\rangle _{BB}\left\langle j+1\right\vert )-\mathrm{H.c.}%
\text{,}  \label{Hm}
\end{eqnarray}%
satisfying $h=$ $h^{\ast }=$ $-h^{\mathrm{T}}$, where basis $\left\{
\left\vert l\right\rangle _{A},\left\vert l\right\rangle _{B},l\in \left[ 1,N%
\right] \right\} \ $is an orthonormal complete set, and $_{A}\langle
l\left\vert l^{\prime }\right\rangle _{B}=\delta _{ll^{\prime }}\delta _{AB}$%
. Matrix $ih $ represents a single-particle representation of a Hermitian
tight-binding ladder.\ Based on the eigenvectors of matrix $h$, the solution
of the Kitaev model with both periodic and open boundary conditions can be
obtained. For a ring system, the same winding number can be obtained from
the eigenstates of $h$. Particularly, the edge modes of $h$\ for an open
chain can be obtained from numerical diagonalization and correspond to the
winding number. In Fig. 2, the plots of the energy levels of $h$\ for the
finite system and the corresponding graphs of the Bloch vector\ in (\ref%
{elllipse})\ demonstrate this bulk-boundary correspondence.

\section{Order parameter}

\label{Order parameter}

It is believed that a superconducting state is a condensation of pairs of
fermions. For the present model, the particle number is not conservative,
and condensation is a consequence of dynamic equilibrium between pair
creation and annihilation. In this section, we introduce an observable to
measure such a condensation.

The phase diagram indicates that there is a threshold value of chemical
potential $\left\vert \mu \right\vert $ for topologically nontrivial phases,
which depends on the factor $\phi $.\ The reason may be that a sufficiently
large $\left\vert \mu \right\vert $\ suppresses the pairing process. On the
other hand, factor\ $\phi $\ also affects this process. We note that the
kinetic energy of a pair of fermions with opposite momenta is $-2\cos
(k+\phi )-2\cos (-k+\phi )$, which is zero for all $k$ when taking $\phi
=\pi /2$, $3\pi /2$. Then, the scattering process of BCS-pair $c_{k}^{\dag
}c_{-k}^{\dag }c_{-k^{\prime }}c_{k^{\prime }}$\ is favorable. This analysis
indicates that the transition rate of BCS-pairs in the nontrivial phases is
relatively larger than that in the trivial phases and depends on the factor $%
\phi $. To quantitatively characterize the pairing process, we introduce the
operator%
\begin{equation}
\hat{O}_{k}=s_{k}^{y}=\frac{i}{2}\left( c_{k}c_{-k}-c_{-k}^{\dag
}c_{k}^{\dag }\right) ,
\end{equation}%
where $\left( s_{k}^{\pm },s_{k}^{y}\right) $\ with%
\begin{equation}
s_{k}^{-}=\left( s_{k}^{+}\right) ^{\dag }=c_{k}c_{-k},
\end{equation}%
are pseudo spin operators. Obviously, for a given state $\left\vert \psi
\right\rangle $, quantity $|\left\langle \psi \right\vert \hat{O}%
_{k}\left\vert \psi \right\rangle |$ measures the rate of transition for a
pair in the $k$ channel and the population of pairs. The corresponding order
parameters are defined by the average magnitude over all channels, i.e.,%
\begin{equation}
O=\frac{2}{N}\sum_{\pi >k>0}|\left\langle \psi \right\vert \hat{O}%
_{k}\left\vert \psi \right\rangle |.
\end{equation}%
In general, $O$ is believed to have different values for different given
states. Nonzero $O$ means that state $\left\vert \psi \right\rangle $ is a
superconducting state. In the following, we derive the analytical
expressions of $O$ for the ground states of the system in different regions
and study their behaviors at the phase boundaries.

For the ground state $\left\vert G\left( \mu ,\phi \right) \right\rangle $
we have

\begin{equation}
O_{\mathrm{g}}\left( \mu ,\phi \right) =\frac{\Delta }{N}\sum\limits_{k>0}%
\frac{\sin k}{r_{k}}.  \label{O}
\end{equation}%
In the limit $N\longrightarrow \infty $, we have $O\left( \mu ,\phi \right)
= $ $\left( \Delta /2\pi \right) \int_{0}^{\pi }\sin k/r_{k}\mathrm{d}k$,
which can be expressed explicitly as 
\begin{equation}
O_{\mathrm{g}}=\frac{1}{\pi a}\left\vert \ln \Gamma \right\vert ,  \label{OO}
\end{equation}%
with%
\begin{equation}
\Gamma =\left\{ 
\begin{array}{cc}
\frac{ia\left\vert \cos \phi \right\vert \Delta +2\cos ^{2}\phi -\Delta ^{2}%
}{\Delta ^{2}} & \left\vert \mu \right\vert <\left\vert \cos \phi \right\vert
\\ 
\frac{2\left\vert \mu \right\vert +i\Delta a}{2\left\vert \mu \right\vert
-i\Delta a} & \text{otherwise}%
\end{array}%
\right. ,
\end{equation}%
where $a=2\sqrt{\Delta ^{2}-\cos ^{2}\phi }/\Delta $.

\begin{figure*}[tbh]
\centering
\includegraphics[width=1.0\textwidth]{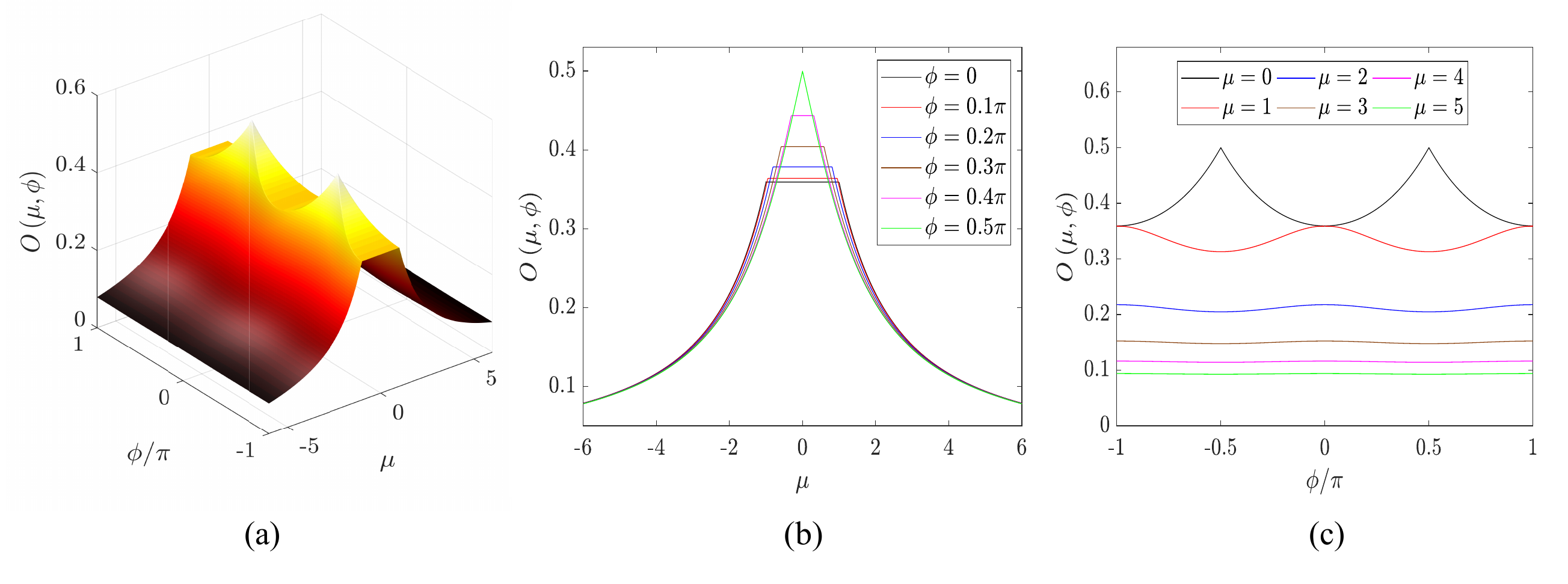}
\caption{(a) Color contour plots of numerical results of order parameter $%
O\left( \protect\mu ,\protect\phi \right) $ defined in (\protect\ref{O}).
(b) Plots of $O\left( \protect\mu ,\protect\phi _{0}\right) $ for several
representative values of $\protect\phi _{0}$. (c) Plots of $O\left( \protect%
\mu _{0},\protect\phi \right) $ for several different $\protect\mu _{0}$.
The parameters are $N=1000$, $J=1$ and $\Delta =1.5$. It is clear that $O$\
is $\protect\mu $\ independent in the nontrivial region and reach its maxima
at the triple critical points.}
\label{fig3}
\end{figure*}

We note that $O_{\mathrm{g}}$\ is\ independent of $\mu $ within the
nontrivial phases. In general, a sudden change in the ground state occurs at
the boundary. Obviously, $O_{\mathrm{g}}$\ is nonanalytic at the phase
boundary and can identify the phase diagram. Notably, $O_{\mathrm{g}}$\
reaches its maxima at the triple critical points $\left( \mu ,\phi \right) =$
$\left( 0,\pm \pi /2\right) $. In Fig. \ref{fig3}, we plot the profiles of $%
O_{\mathrm{g}}$ in the $\mu $-$\phi \ $plane. This clearly demonstrates the
above points. 

\section{Pair condensation with ODLRO}

\label{Pair condensation with ODLRO}

In this section, we explore the implication of the obtained result. First,
we would like to point out that the BCS-pair correlation is intimately
related to the pair in real space due to the following relation in a large $%
N $ limit

\begin{equation}
s^{y}=-\frac{1}{\pi }\sum_{l}\left( \zeta _{l}^{\dag }+\zeta _{l}\right) ,
\end{equation}%
where we introduce a local pair operator%
\begin{equation}
\zeta _{l}=\sum\limits_{\text{\textrm{odd} }r>0}\frac{1}{r}c_{l}c_{l+r}.
\end{equation}%
Second, the maximum of $O_{\mathrm{g}}=1/2$\ is also the maximum of the
operator $\sum\nolimits_{k>0}\hat{O}_{k}$. This indicates that the
corresponding ground state is the eigenstate of $\sum\nolimits_{k>0}\hat{O}%
_{k}$\ and that the hopping term has no contribution to the energy.\ A
similar case occurs in the $\eta $-pairing state in the Hubbard model, where
the single-particle scattering is suppressed by\ coherent destructive
interference\ for the states with ODLRO.\ We note that the operator $s^{+}$\
represents the local pairing creation, which is similar to the $\eta $-pair
operator situation with $N\longrightarrow \infty $.\ These results motivate
us to investigate the ODLRO\ in the ground state of the present model.

We start with the case with $\mu =0$\ and $\phi =\pi /2$, where the
Hamiltonian becomes $H=H_{\mathrm{T}}+H_{\mathrm{P}}$, with%
\begin{eqnarray}
H_{\mathrm{T}} &=&2\sum\limits_{\pi >k>0}\sin k\left( c_{k}^{\dag
}c_{k}-c_{-k}^{\dag }c_{-k}\right) , \\
H_{\mathrm{P}} &=&2i\Delta \sum\limits_{\pi >k>0}\sin k\left( c_{-k}^{\dag
}c_{k}^{\dag }+c_{-k}c_{k}\right) .
\end{eqnarray}%
It is easy to check that, the hopping term is decoupled from the pairing
term, i.e.,%
\begin{equation}
\left[ H_{\mathrm{T}},H_{\mathrm{P}}\right] =0.
\end{equation}%
One can divide the set of $k\in \left( 0,\pi \right) $ into two subsets $%
\left\{ k_{a}\right\} $\ and $\left\{ k_{b}\right\} $ arbitrarily, $\left\{
k\right\} =\left\{ k_{a}\right\} \oplus \left\{ k_{b}\right\} $. A set of
eigenstates can be constructed in the form%
\begin{equation}
\left\vert \Psi \right\rangle =\underset{k\in \left\{ k_{a}\right\} }{\prod }%
c_{\pm k}^{\dag }\left\vert 0\right\rangle _{k}\left\vert 0\right\rangle
_{-k}\underset{k\in \left\{ k_{b}\right\} }{\prod }\frac{i\pm c_{-k}^{\dag
}c_{k}^{\dag }}{\sqrt{2}}\left\vert 0\right\rangle _{k}\left\vert
0\right\rangle _{-k}.
\end{equation}%
with the eigen energy

\begin{equation}
E=\pm 2\sum\limits_{k\in \left\{ k_{a}\right\} }\sin k\mp 2\Delta
\sum\limits_{k\in \left\{ k_{b}\right\} }\sin k.
\end{equation}%
We note that under the condition $\Delta >J>0$, the ground state is%
\begin{equation}
\left\vert G\right\rangle =\prod\limits_{k}\frac{i+c_{-k}^{\dag }c_{k}^{\dag
}}{\sqrt{2}}\left\vert 0\right\rangle _{k}\left\vert 0\right\rangle _{-k},
\label{G}
\end{equation}%
which is also the eigenstate of the operator $\sum\nolimits_{k>0}\hat{O}_{k}$%
.

Importantly, it can be rewritten in the form%
\begin{equation}
\left\vert G\right\rangle =\sum_{n=0}^{N/2}d_{n}\left\vert \psi
_{n}\right\rangle ,
\end{equation}%
where the coefficient%
\begin{equation}
d_{n}=\sqrt{C_{N/2}^{n}}\sin ^{n}\left( \pi /4\right) \left( i\cos \left(
\pi /4\right) \right) ^{N/2-n},
\end{equation}%
and the basis 
\begin{equation}
\left\vert \psi _{n}\right\rangle =\frac{1}{\Omega _{n}}(s^{+})^{n}\left%
\vert 0\right\rangle ,\left\vert 0\right\rangle =\underset{k>0}{\prod }%
\left\vert 0\right\rangle _{k}\left\vert 0\right\rangle _{-k},
\end{equation}%
with the normalization factor $\Omega _{n}=\left( n!\right) \sqrt{C_{N/2}^{n}%
}$. It is clear that state\ $\left\vert G\right\rangle $\ is a coherent
state of zero-momentum-pair condensation. In fact, direct derivation shows
that 
\begin{equation}
\left\langle G\right\vert s^{+}s^{-}\left\vert G\right\rangle =\frac{N(N+2)}{%
16}.
\end{equation}%
In the limit $N\longrightarrow \infty $, $\left\langle G\right\vert
s^{+}s^{-}\left\vert G\right\rangle $\ is contributed by $N^{2}$\ terms such
as $\left( 2/\pi \right) ^{2}\zeta _{l}^{\dag }\zeta _{l^{\prime }}$. It is
presumably that the correlator%
\begin{equation}
C_{ll^{\prime }}=\left\langle G\right\vert \zeta _{l}^{\dag }\zeta
_{l^{\prime }}\left\vert G\right\rangle ,
\end{equation}%
is finite when the distance $\left\vert l-l^{^{\prime }}\right\vert \gg 1$.
The direct derivation of $\zeta _{l}\left\vert G\right\rangle $\ in the
Appendix shows that 
\begin{equation}
C_{ll^{\prime }}=\frac{\pi ^{2}}{64},\text{for }\left\vert l-l^{^{\prime
}}\right\vert \gg 1.  \label{Cll}
\end{equation}%
Then, we conclude that the ground state at the triple critical point
possesses the exact ODLRO.

\section{Summary}

\label{Summary}

We have presented a finding that the ground state of the Kitaev model can
possess ODLRO in the presence of a uniform phase gradient on the pairing
term. The phase gradient plays a different role from the chemical potential
in influencing the phase diagram no matter what kind of boundary condition
is considered. Accordingly, the Majorana edge modes can be controlled by the
phase gradient,\textbf{\ }providing an alternative avenue for topological
quantum computation. In addition, it has been shown that the BCS-pair order
parameter is chemical potential $\mu $ independent but phase gradient $\phi $%
\ dependent within the topologically nontrivial phase. At the triple
critical point, the ground state is the condensation of zero-momentum
fermion pairs with different pair densities. In this regard, it is an analog
of the $\eta $-pairing state\ in the Hubbard model, replacing the on-site
single pair with a local $p$-wave pair. This study provides insight into the
long-range order that emerges from the interplay among the single-particle
kinetic energy, chemical potential, and phase gradient in the 1D Kitaev
model.

\acknowledgments This work was supported by the National Natural Science
Foundation of China (under Grant No. 11874225).

\section*{Appendix}

\appendix\setcounter{equation}{0} \renewcommand{\theequation}{A%
\arabic{equation}} \setcounter{figure}{0} \renewcommand{\thefigure}{A%
\arabic{figure}}

In this Appendix, the derivation of the correlator in Eq. (\ref{Cll}) is
presented. By Fourier transformation, we have

\begin{eqnarray}
\zeta _{l} &=&\sum\limits_{\text{\textrm{odd} }r>0}\frac{1}{r}c_{l}c_{l+r} 
\notag \\
&=&\frac{1}{N}\sum\limits_{\text{\textrm{odd} }r>0}\sum_{k,k^{\prime }}\frac{%
1}{r}e^{i\left( k+k^{\prime }\right) l}e^{ik^{\prime }r}c_{k}c_{k^{\prime }},
\end{eqnarray}%
and%
\begin{equation}
\zeta _{l^{\prime }}^{\dag }=\frac{1}{N}\sum\limits_{\text{\textrm{odd} }%
r>0}\sum_{k,k^{\prime }}\frac{1}{r}e^{-i\left( k+k^{\prime }\right)
l^{\prime }}e^{-ik^{\prime }r}c_{k^{\prime }}^{\dag }c_{k}^{\dag }.
\end{equation}%
The correlator has the form

\begin{eqnarray}
\zeta _{l^{^{\prime }}}^{\dag }\zeta _{l} &=&\frac{1}{N^{2}}%
\sum_{k_{1},k_{2},k_{3},k_{4}}\rho _{-k_{1}}\rho _{k_{4}}e^{-i\left(
k_{2}+k_{1}\right) l^{\prime }}e^{i\left( k_{3}+k_{4}\right) l}  \notag \\
&&\times c_{k_{1}}^{\dag }c_{k_{2}}^{\dag }c_{k_{3}}c_{k_{4}},
\end{eqnarray}%
where the coefficient%
\begin{equation}
\rho _{k}=\sum\limits_{\text{\textrm{odd} }r>0}^{\infty }\frac{1}{r}e^{ikr}=%
\frac{i\pi }{4}+\frac{1}{2}\ln \cot \left( k/2\right) ,
\end{equation}%
and specifically

\begin{equation}
\rho _{k}-\rho _{-k}=\pm \frac{i\pi }{2}.
\end{equation}

Based on the expectation value of four operators

\begin{eqnarray}
&&\left\langle G\right\vert c_{k_{1}}^{\dag }c_{k_{2}}^{\dag
}c_{k_{3}}c_{k_{4}}\left\vert G\right\rangle  \notag \\
&=&\frac{1}{4}\left\{ 
\begin{array}{c}
\delta _{k_{2},k_{3}}\delta _{k_{1},k_{4}},\left( \left\vert
k_{2}\right\vert >\left\vert k_{1}\right\vert ,\left\vert k_{3}\right\vert
>\left\vert k_{4}\right\vert \right) \\ 
\delta _{k_{2},k_{3}}\delta _{k_{1},k_{4}},\left( \left\vert
k_{2}\right\vert <\left\vert k_{1}\right\vert ,\left\vert k_{3}\right\vert
<\left\vert k_{4}\right\vert \right) \\ 
-\delta _{k_{2},k_{4}}\delta _{k_{1},k_{3}},\left( \left\vert
k_{2}\right\vert <\left\vert k_{1}\right\vert ,\left\vert k_{3}\right\vert
>\left\vert k_{4}\right\vert \right) \\ 
-\delta _{k_{1},k_{3}}\delta _{k_{2},k_{4}},\left( \left\vert
k_{2}\right\vert >\left\vert k_{1}\right\vert ,\left\vert k_{3}\right\vert
<\left\vert k_{4}\right\vert \right) \\ 
2\delta _{k_{2},k_{3}}\delta _{k_{1},k_{4}}\delta _{k_{2},-k_{1}}, \\ 
-2\delta _{k_{1},k_{3}}\delta _{k_{2},k_{4}}\delta _{k_{4},-k_{1}}, \\ 
-\mathrm{sign}(k_{1}k_{3})\delta _{k_{2},-k_{1}}\delta
_{k_{4},-k_{3}},\left( \left\vert k_{3}\right\vert \neq \left\vert
k_{1}\right\vert \right)%
\end{array}%
\right. ,
\end{eqnarray}

we have%
\begin{eqnarray}
&&\left\langle G\right\vert \zeta _{l^{^{\prime }}}^{\dag }\zeta
_{l}\left\vert G\right\rangle  \notag \\
&=&\frac{1}{4N^{2}}\sum_{\left\vert k_{2}\right\vert \neq \left\vert
k_{1}\right\vert }\left( \rho _{-k_{1}}\rho _{k_{1}}-\rho _{-k_{1}}\rho
_{k_{2}}\right) e^{i\left( k_{1}+k_{2}\right) \left( l-l^{\prime }\right) } 
\notag \\
&&+\frac{1}{2N^{2}}\sum_{k_{1}}\left( \rho _{-k_{1}}\rho _{k_{1}}-\rho
_{-k_{1}}\rho _{-k_{1}}\right)  \notag \\
&&-\frac{1}{4N^{2}}\sum_{\left\vert k_{3}\right\vert \neq \left\vert
k_{1}\right\vert }\mathrm{sign}(k_{1}k_{3})\rho _{-k_{1}}\rho _{-k_{3}}.
\end{eqnarray}

Three terms can be further expressed explicitly in the form

\begin{eqnarray}
&&\sum_{\left\vert k_{2}\right\vert \neq \left\vert k_{1}\right\vert }\left(
\rho _{-k_{1}}\rho _{k_{1}}-\rho _{-k_{1}}\rho _{k_{2}}\right) e^{i\left(
k_{1}+k_{2}\right) \left( l-l^{\prime }\right) }  \notag \\
&=&\sum_{k_{1}>0,k_{2}>k_{1}}2\{\left\vert \rho _{k_{1}}-\rho
_{k_{2}}\right\vert ^{2}\cos \left[ \left( k_{2}+k_{1}\right) \left(
l-l^{\prime }\right) \right]  \notag \\
&&+\left\vert \rho _{k_{2}}-\rho _{-k_{1}}\right\vert ^{2}\cos \left[ \left(
k_{2}-k_{1}\right) \left( l-l^{\prime }\right) \right] \},
\end{eqnarray}

and

\begin{eqnarray}
&&\sum_{k_{1}}\left( \rho _{-k_{1}}\rho _{k_{1}}-\rho _{-k_{1}}\rho
_{-k_{1}}\right)  \notag \\
&=&\sum_{k_{1}>0}\left( \rho _{k_{1}}-\rho _{-k_{1}}\right) \left( \rho
_{-k_{1}}-\rho _{k_{1}}\right) =\sum_{k_{1}>0}\frac{\pi ^{2}}{4},
\end{eqnarray}

and

\begin{eqnarray}
&&\sum_{\left\vert k_{3}\right\vert \neq \left\vert k_{1}\right\vert }%
\mathrm{sign}(k_{1}k_{3})\rho _{-k_{1}}\rho _{-k_{3}}  \notag \\
&=&\sum_{k_{1},k_{3}>0,k_{3}\neq k_{1}}-\left( \rho _{k_{1}}-\rho
_{-k_{1}}\right) \left( \rho _{-k_{3}}-\rho _{k_{3}}\right)   \notag \\
&=&\sum_{k_{1},k_{3}>0,k_{3}\neq k_{1}}-\frac{\pi ^{2}}{4}.
\end{eqnarray}

Therefore, we have

\begin{eqnarray}
&&\left\langle G\right\vert \zeta _{l^{^{\prime }}}^{\dag }\zeta
_{l}\left\vert G\right\rangle  \notag \\
&=&\frac{1}{8\pi ^{2}}\int_{0}^{\pi }\int_{0}^{k_{2}}\{\left\vert \rho
_{k_{1}}-\rho _{k_{2}}\right\vert ^{2}\cos \left[ \left( k_{2}+k_{1}\right)
\left( l-l^{\prime }\right) \right]  \notag \\
&&+\left\vert \rho _{k_{2}}-\rho _{-k_{1}}\right\vert ^{2}\cos \left[ \left(
k_{2}-k_{1}\right) \left( l-l^{\prime }\right) \right] \}\mathrm{d}k_{1}%
\mathrm{d}k_{2}  \notag \\
&&+\frac{\pi ^{2}}{64}.
\end{eqnarray}

We note that two terms

\begin{equation}
\left\vert \rho _{k_{1}}-\rho _{k_{2}}\right\vert ^{2}=\frac{1}{4}\left\vert
\ln \frac{\cot (k_{1}/2)}{\cot (k_{2}/2)}\right\vert ^{2},
\end{equation}%
and%
\begin{equation}
\left\vert \rho _{k_{2}}-\rho _{-k_{1}}\right\vert ^{2}=\frac{1}{4}%
\left\vert i\pi +\ln \frac{\cot (k_{2}/2)}{\cot (k_{1}/2)}\right\vert ^{2},
\end{equation}%
are finite except at the boundary, and cannot induce the divergence of the
integral. In the case $\left\vert l-l^{\prime }\right\vert \gg 1$, the high
frequency oscillation of terms $\cos \left[ \left( k_{2}+k_{1}\right) \left(
l-l^{\prime }\right) \right] $ and $\cos \left[ \left( k_{2}-k_{1}\right)
\left( l-l^{\prime }\right) \right] $ result in the vanishing contribution.
Therefore, we obtain the conclusion

\begin{equation}
\left\langle G\right\vert \zeta _{l^{^{\prime }}}^{\dag }\zeta
_{l}\left\vert G\right\rangle =\frac{\pi ^{2}}{64}.
\end{equation}

\end{document}